\documentstyle[sprocl]{article}
\begin{document}
	\title{EXACT RENORMALIZATION GROUP WITH FERMIONS}
	\author{J. COMELLAS}
	\address{Departament d'Estructura i Constituents de la Mat\`{e}ria\\
		Facultat de F\'{\i}sica, Universitat de Barcelona\\
		Diagonal 647, 08028 Barcelona, Spain}
	\maketitle\abstracts{The development of the Exact Renormalization Group
		for fermionic theories is presented, together with its application
		to the chiral Gross-Neveu model.  We focus on the reliability
		of various approximations, specifically
		the derivative expansion and further
		truncations in the number of fields.  The main differences with
		bosonic theories are discussed.}
	\section{Introduction}
		The set of ideas enclosed in the
		Renormalization Group (RG, hereafter)
		has led to a
		variety of developments in many fields, as this Conference has made
		apparent.
		From a particle theorist point of view they englobe a bunch of ideas
		form which we may understand what a quantum field theory is.  Moreover,
		it provides a framework for nonperturbative calculations.
		
		In recent years, there has been an intensive development of the field
		mainly for scalar theories.\cite{yuri}
		It can probably be said that we thoroughly understand all the
		subtleties the RG reserves for us in this case.

		Nevertheless what would ultimately justify the whole approach,
		as applied for Particle Physics,
		is the construction of Lorentz and gauge invariant nonperturbative
		equations, manageable for reliable approximations.  The hope of
		RG practitioners is that
		we are really not that
		far from there.\footnote{We have to mention that although the search
			for nonperturbative gauge invariant equations is still open,
			there are a variety of results for perturbative definitions of
			a gauge invariant theory based on these grounds.\cite{marisa}}
		
		Particle physicists may imagine themselves, thus,
		applying in the nearby future the
		powerfulness of the approach to atack long-standing nonperturbative
		problems for, say, quantum gluodynamics.  And the next
		step would probably be the introduction of matter to have the full
		physical theory.

		At this point it will for sure be helpful to have already understood
		the peculiarities associated to fermions on their own, both conceptual
		and technical.  This is the reason we believe that we should try,
		as it has been done for bosons, to master as deeply
		as we can fermionic theories, even with no additional fields.

		We do not want to suggest that the features of fermionic equations have
		to be significantly different from the ones for bosons.  On the contrary,
		we believe that they have to be ultimately a direct translation
		of ideas from
		one subject to the other.
		What we do mean is that these
		features may not be noticeable in any obvious way.  On the other hand,
		new technicalities may also appear.  And
		before embarking ourselves in a
		more ambitious project these peculiarities must be previously
		worked out and clearly understood.

		We thus want to present some recent work in that direction.\cite{ours}
		Namely, the study of a two dimensional sample model (the so-called
		Gross-Neveu model$\,$\cite{gn})
		which we will define below.  This
		model is sufficiently simple in order to be able to carry out the algebra
		as far as we need but that
		it still captures the essentials of the approach
		for the Grassman case.
		
		Let us briefly review the main ideas involved
		in bosonic theories as they are studied by Polchinski.\cite{polchinski}
		We first have to choose a regulator adequated for our purposes.
		It is done by simply
		modifying the propagator
			\( P(p)
			\)
		\begin{equation}
			P(p)=\frac{1}{p^2}
		\end{equation}
		to
		\begin{equation}
			P_{\Lambda}(p)=\frac{K(p^2/\Lambda^{2})}{p^2}
		\end{equation}
		where
			\( \Lambda
			\)
		is a momentum-space cutoff and
			\( K(x)
			\)
		a regulating function which decays sufficiently
		rapid to zero when
			\( x\rightarrow\infty
			\).

		With this kind of regulator, a quick (and somehow
		sloppy) argument that leads
		to an appropriate RG equation is to identify all the
			\( \Lambda
			\)-dependences in a partly integrated action by signaling
		all possible occurrences of the propagator and multiplying by the
			\( \Lambda
			\)-derivative of it.  In this manner we immediately obtain
		\begin{equation}
		\label{RGeq:bos}
			-\Lambda\frac{d}{d\Lambda}S_{int}
			\equiv\dot{S}_{int}
			=\frac{1}{2}\frac{\delta S_{int}}{\delta \phi}\cdot\dot{P}_{\Lambda}
				\cdot\frac{\delta S_{int}}{\delta \phi}
			-\frac{1}{2}\mbox{tr}
				\left(
					\dot{P}_{\Lambda}\cdot\frac{\delta^{2}S_{int}}{\delta
					\phi\delta\phi}
				\right)
		\end{equation}
		with
		\begin{equation}
			S_{int}=S-\frac{1}{2}\phi\cdot P_{\Lambda}^{-1}\cdot\phi
		\end{equation}
		and $S$ the full action.
		The first term takes into account tree-type propagators and the second
		one loop-type propagators.
		We are using a compact notation regarding the propagator as a matrix with
		a dot standing for matrix multiplication.

		The equation for a pure fermionic theory can be written in a similar form.
		From the propagator
		\begin{equation}
			P_{\Lambda}=i\not\!{p}\,\frac{K(p^{2}/\Lambda^{2})}{p^{2}}
		\end{equation}
		we obtain the RG equation
		\begin{equation}
			\dot{S}_{int}
			=\frac{\delta S_{int}}{\delta \psi}\cdot\dot{P}_{\Lambda}\cdot
				\frac{\delta S_{int}}{\delta\bar{\psi}}
			-\mbox{tr}
				\left(
					\frac{\delta}{\delta\psi}\cdot\dot{P}_{\Lambda}\cdot
					\frac{\delta S_{int}}{\delta\bar{\psi}}
				\right)
		\end{equation}
		
		Returning to
			\( S
			\)
		and expressing the resultant equation in dimensionless variables
		it is obtained
		\begin{eqnarray}
		\label{RGeq:fer}
			\dot{S}
			&=&2K'(p^{2})\frac{\delta S}{\delta\psi}\cdot i\not\!{p}
				\cdot\frac{\delta S}{\delta\bar{\psi}}
			-\mbox{tr}
				\left(
					2K'(p^{2})\frac{\delta}{\delta\psi}\cdot i\not\!{p}\,
					\frac{\delta S}{\delta\bar{\psi}}
				\right)\nonumber
			\\&-&2p^{2}\frac{K'(p^{2})}{K(p^{2})}
				\left(
					\bar{\psi}\cdot\frac{\delta S}{\delta \bar{\psi}}
					+\psi\frac{\delta S}{\delta \psi}
				\right)
			\\&+&dS\nonumber
			\\&+&\frac{1-d+\eta(t)}{2}
				\left(\bar{\psi}\frac{\delta S}{\delta\bar{\psi}}
					+\psi\frac{\delta S}{\delta\psi}
				\right)
			-\left(
				\bar{\psi}\cdot p^{\mu}\frac{\partial'}{\partial p^{\mu}}
					\frac{\delta S}{\delta\bar{\psi}}
				+\psi\cdot p^{\mu}\frac{\partial'}{\partial p^{\mu}}
					\frac{\delta S}{\delta \psi}
			\right)\nonumber
		\end{eqnarray}
		where we work on a $d$-dimensional Euclidean space;
			\( \eta
			\)
		is the anomalous dimension (needed to obtain a physically
		interesting fixed
		point);
			\( t\equiv-\ln \Lambda
			\);
		and the prime in
			\( \frac{\partial'}{\partial p^{\mu}}
			\)
		means that the derivative does not act on the momentum conservation
		delta functions and thus only serves to count powers of momenta.

		Note the first difference between bosons and fermions.  Due to the
		different structure of the propagators, the fermionic equation presents
		an explicit factor of $p$ in the first two terms of the right hand
		side of the RG equation~\ref{RGeq:fer}
		while this is not the case for bosons (Eq.~\ref{RGeq:bos}).
		This may just look
		like a technical
		remark without any relevance.  However it turns out that the
		seemingly most powerful approximations to these equations are based
		on the so-called derivative expansion\cite{tim}
		whose first order term is obtained by restricting the action to
		be a kinetic term plus a general potential term with no derivatives.
		In a fermionic theory this approximation will not be feasible, because
		we will be left only with a fairly simple linear equation.
		The derivative
		expansion should nevertheless be applicable, but it would lead to more
		complicated structures even at first non-trivial order.
	\section{The model}
		Let us now apply the RG equation~\ref{RGeq:fer}
		to the so-called chiral Gross-Neveu model.\cite{gn}

		As any other field theory,
		it is best defined through its
		symmetries.  We will consider thus
		an Euclidean
		invariant $N$-flavoured model,
		with an
			\( U(N)\times U(N)
			\)
		internal symmetry group.  It is also chosen to obey
		the discrete symmetries of parity, charge conjugation and reflection
		hermiticity.\cite{zinn-justin}
		
		When one imposes these restictions and further use Fierz reorderings,
		it is easily sown that there appear only three basic structures,
		\begin{eqnarray}
			V_{12}^{j}&\equiv&
				\bar{\psi}^{a}(p_{1})\gamma^{j}\psi^{a}(p_{2})\nonumber\\
			S_{12}S_{34}-P_{12}P_{34}&\equiv&
				\bar{\psi}^{a}(p_1)\psi^{a}(p_2)\bar{\psi}^{b}(p_3)\psi^{b}(p_4)
					\nonumber\\
				&-& \bar{\psi}^{a}(p_1)\gamma_{s}\psi^{a}(p_2)
					\bar{\psi}^{b}(p_3)\gamma_{s}\psi^{b}(p_4)\\
			S_{12}P_{34}-P_{12}S_{34}&\equiv&
				\bar{\psi}^{a}(p_1)\psi^{a}(p_2)
					\bar{\psi}^{b}(p_3)\gamma_{s}\psi^{b}(p_4)
					\nonumber\\
				&-& \bar{\psi}^{a}(p_1)\gamma_{s}\psi^{a}(p_2)
					\bar{\psi}^{b}(p_3)\psi^{b}(p_4)\nonumber
		\end{eqnarray}
		which
		have to be combined in an arbitrary way with combinations of
		momenta.

		The next step is to define a reasonable approximation to handle the
		above functional-derivative equation.
		We would like to choose one that closely resembles the bosonic derivative
		approximation.  Nevertheless,
		due to the number of different structures it is not that easy to
		parametrize the general action up to, say, two derivatives
		while maintaining
		arbitrary the number of fields.  Moreover, we should keep in mind that
		the allowed action is, as long as we are working with
		a finite number of different
		species,
		composed by a finite number of operators:\ the Grassman character
		of our variables constraints the number of fields allowed at one
		point of space.

		We have already commented that derivative terms
		should also be included.  In fact, this is an
		important point because in
			\( d=2
			\)
		the anomalous dimension
			\( \eta
			\)
		usually plays an important
		role: we would probably be too naive if we try to obtain numbers without
		letting it to be nonzero.
		In fact two derivatives may seem to do the job.  However, once one
		goes through the calculations, it turns out to be quite clear that
			\( \eta =0
			\)
		is the only consistent value.
		This implies that we need at least three derivatives.

		The maximum number of fields was chosen to be six.  This seems a number
		both sufficiently low in order to keep the action relatively simple
		and sufficiently high to let non-trivial results appear.

		The action thus obtained has the usual kinetic term; one term with
		three derivatives and only two fields; two derivative-free four-fields
		operators
		\begin{equation}
			g_{1}(S_{12}S_{34}-P_{12}P_{34})\ ,\ \ \ \ \ g_{2}V_{12}^{j}V_{34}^{j}
		\end{equation}
		with coupling constants
			\( g_{1}
			\)
		and
			\( g_{2}
			\);
		eleven operators with also four fields but two derivatives; and
		ninety-two six-fermions operators, five of them
		with only one derivative and
		the rest with three derivatives.

		After some algebra we can now obtain the set of beta functions.
		The fixed points are the solutions for these functions to vanish.
		They are a set of 106 non-linear algebraic equations.

		Up to this point, the function
			\( K(p^{2})
			\)
		can be mainted arbitrary, thus keeping some freedom of chosing
		a scheme.  The fixed point solutions in our
		approximation will in general depend on two
		parameters which serve as a scheme parametrization.
		In principle,
		this should not worry us, because it is well known that the actual
		expression of the fixed point action has no intrinsic physical
		meaning.

		For Particle Physics it is specially interesting the
		value of the relevant directions from the fixed points.
		That is, we linearize the RG transformations,
		\begin{equation}
			\dot{g}_{i}={\cal R}_{i}(g_{j})
		\end{equation}
		to
		\begin{equation}
			\dot{g}_{i}={\cal R}_{ij}\cdot\delta g_{j}\ ,\ \ \ \ \
			{\cal R}_{ij}\equiv
				\left.\frac{\partial{\cal R}_{i}}{\partial g_{j}}\right|_{g^{0}}
		\end{equation}
		where
			\( g^{0}
			\)
		is the fixed-point solution and
			\( \delta g_{j}
			\)
			are the deviations from it.
		The number of positive eigenvalues of the matrix
			\( {\cal R}_{ij}
			\)
		coincide with the number of possible parameters we can fine-tune in the
		corresponding cutoff-free theory$\,$\cite{wilson}
		and the actual value of these eigenvalues gives the speed of departure
		from the fixed point.

		These eigenvalues are directly related to the so-called critical
		exponents in the terminology of second-order phase transitions.
		They are universal and, therefore, they should be
		free from schemes dependences.
		In our approximation, however,
		this is not so, as often happens with truncations.
		The scheme ambiguities are solved by a translation of the principle
		of minimal sensitivity used in perturbative
		calculations.\cite{stevenson}
	\section{Results}
		We now sketch the main results.

		The equations simplify enormously when
			\( N\rightarrow\infty.
			\)
		Two fixed points can be clearly identified.  One of them with vanishing
		anomalous dimension (it is of order
			\( N^{-1}
			\)) and
		with the most relevant eigenvalue
			\( \sqrt{17} - 3
			\)
		in this approximation.  Moreover the coupling constant
			\( g_{2}
			\)
		which corresponds to
			\( U(1)
			\)
		Thirring-like
		excitations becomes free (we have, in fact, a line of fixed points) and
			\( g_{1}
			\)
		is also of order
			\( N^{-1}
			\).
		All these features but the anomalous dimension remind the fixed point
		solution found by Dashen and Frishman.\cite{df}
		
		The other solution, which corresponds to a different definition of
		the large $N$ limit (different assumed
		$N$ dependences of the coupling constants) has a non-vanishing anomalous
		dimension.  It is scheme dependent with a range of variation of
		1.11--1.14 for most of the schemes.  The most relevant eigenvalue
		is also scheme-dependent with a range of 2.1--2.3 and, unlike the previous
		case, there are no free parameters.

		Before going on we must comment on a quite unpleasant feature of this kind
		of approximations.  By now it is generally believe that any approximation
		based on truncations leads to a system of fixed-point equations with
		many spurious solutions.\cite{spur}
		It seems that a pure derivative
		expansion (that is, one with a truncation in the number of derivatives
		but without any further truncation in the number of fields) cures
		all this kind of problems.
		We work with a truncated action and thus we expect on general grounds
		that this unwanted peculiarity appears and, actually, it does.
		The solution of the puzzle is not always simple.  One usually tries
		to discriminate among solutions by checking the stability of the obtained
		ones either going one step beyond in the approximation or else
		tuning some parameters.  In our case we are lucky to have nitid
		results in the large $N$ limit.  Therefore,
		we take as reliable solutions only those whose limit when
			\( N\rightarrow\infty
			\)
		concides with one of the solutions found above.

		This procedure will probably not be available in all cases and one should
		wonder if there is any systematic procedure to deal with the problem
		without relying on technical details of the studied
		model.
		Of course, one can always try to perform a true derivative expansion
		instead of mutilating each term as we have done.  It should
		eliminate at once the spurious results.  In fact we have a special
		case, which we will refer later on that suggests that this is true.
		Nevertheless the expansion proposed is not that simple, specially for
		$N$ moderately large.  Moreover, it seems to be difficult to deal
		with different values of $N$ simultaneously and still preserving
		each term in the expansion without truncating it at some arbitrary
		point.

		One solution for finite $N$ matches
		the first one
		discussed above, with the most relevant eigenvalue smoothly decreasing
		to
			\( \sqrt{17} - 3
			\)
		and with
			\( N\cdot\eta
			\)
		increasing with N to 4.87\ldots\@\ Unlike the strict large $N$ limit,
		we do not find, nevertheless, a line of fixed points but an isolated
		one.  We blame this feature to the crudeness of the approximation.

		The solution that matches the second one above has a much more conspicuous
		behaviour.  In fact it is valid only for
			\( N>142.8
			\).
		At this value it matches another branch of solutions, which
		exists even when
			\( N\rightarrow\infty
			\)
		although the couplings do not scale with $N$ as integer powers but as
		noninteger ones.  Both the anomalous dimension and the most relevant
		eigenvalue present strong scheme dependences, quite difficult to
		disentangle.

		Finally, we can consider separately the
			\( N=1
			\)
		case.  It is worth going through it because it is a simple case where
		we can treat the equations in a purely derivative approximation:
		due to Fermi statistics, we cannot have more than six fields
		if we consider terms up to three derivatives.  Our action is thus exact
		in this sense.  Also, we have to work out further relations imposed
		by Fierz reorderings, not present for
			\( N\not=1
			\).
		Our action is, therefore,
		even shorter than that.  The results are nevertheless
		very messy and, probably, not reliable.  When considering terms up
		to two derivatives, we find a line of fixed points (as it is expected
		in the Thirring model) with
			\( \eta=0
			\)
		as stated previously.  But once we go to three derivatives,
			\( \eta\not=0
			\) but, unexpectedly, the line of fixed points disappears and we
		obtain only an isolated one.  Nevertheless one piece of nice news comes
		out: the spurious solutions disappear, as expected.

		A final comment is in order.  We have found
			\( \eta
			\)
		by imposing that the normalization of the kinetic term is fixed at
		some standard value.  This is surely not the most general way to
		proceed.  If an exact computation is performed we know that this
		normalization does not matter and we will be able to fix it safely
		to whatever value we want: we will find a whole line of physically
		equivalent fixed points.  It is generally known that this
		kind of symmetry is broken for
		most of the approximations$\,$\cite{bell} (in particular it is broken by
		the derivative expansion).  Moreover, we generally expect that the true
		physical fixed point mixes with non-local ones with similar behaviour
		for the truncated action but with different anomalous dimensions.
		To pick up the local solution
		among the non-local ones, one should try to find the
		reminiscence of the line of fixed points:\
		a marginal redundant operator.
		Its presence would signal that our scheme is truly approximating
		the local fixed point behaviour and not something else.  This would
		probably fixed some, or perhaps even all, of the scheme dependences.
		This analysis has not been performed.

		Summarizing, we have presented
		a fermionic RG equation and an example of its
		application.  It seems that, although technically harder to work with,
		there emerges the same patterns as in the bosonic case.  In particular
		the annoying issue of spurious solutions is also present.
		However, it does seem that with sufficiently accurate work and
		restricting oneself to a derivative expansion without further
		truncations reliable non-trivial results should come out.
	\section*{Acknowledgments}
		I would like to acknowledge J.I. Latorre and
		Tim R. Morris for discussions on this
		and related subjects and A. Travesset
		for reading the manuscript.
		This work has been supported by funds form MEC
		under contract AEN95-0590.
	
\end{document}